%% file: main.tex
\RequirePackage[2020-02-02]{latexrelease}
\documentclass[runningheads]{llncs}
\usepackage{import}
\usepackage{preamble}
\usepackage{algorithm}
\usepackage{algpseudocode}
\usepackage{booktabs}
\usepackage[cal=boondox]{mathalfa}
\usepackage{ stmaryrd }
\usepackage{adjustbox}
\let\svthefootnote\thefootnote
\newcommand\freefootnote[1]{%
  \let\thefootnote\relax%
  \footnotetext{#1}%
  \let\thefootnote\svthefootnote%
}

\usepackage{array}
\newcolumntype{P}[1]{>{\centering\arraybackslash}p{#1}}

\algnewcommand\algorithmicforeach{\textbf{for each}}
\algdef{S}[FOR]{ForEach}[1]{\algorithmicforeach\ #1\ \algorithmicdo}
\algrenewcommand\algorithmicrequire{\textbf{Input:}}
\algrenewcommand\algorithmicensure{\textbf{Output:}}
\begin{document}
%
%\title{Contribution Title\thanks{Supported by organization x.}}
%
\title{CombiGCN: An effective GCN model for Recommender System}

\author{Loc Tan Nguyen\orcidID{0009-0002-4926-398X} \and
Tin T. Tran\orcidID{0000-0003-4252-6898} }
\authorrunning{Loc Tan Nguyen and Tin T. Tran}
%\authorrunning{Short author list}% Part of LEFT running header
\titlerunning{CombiGCN: An effective GCN model for Recommender System}% Part of RIGHT running header
\institute{Faculty of Information Technology, Ton Duc Thang University, Ho Chi Minh city, Vietnam\\
\email{51900375@student.tdtu.edu.vn, trantrungtin@tdtu.edu.vn} }
\maketitle              % typeset the header of the contribution
\freefootnote{* Corresponding author: Tin T. Tran, trantrungtin@tdtu.edu.vn}
  
\begin{abstract}
Graph Neural Networks (GNNs) have opened up a potential line of research for collaborative filtering (CF). The key power of GNNs is based on injecting collaborative signal into user and item embeddings which will contain information about user-item interactions after that. However, there are still some unsatisfactory points for a CF model that GNNs could have done better. The way in which the collaborative signal are extracted through an implicit feedback matrix that is essentially built on top of the message-passing architecture of GNNs, and it only helps to update the embedding based on the value of the items (or users) embeddings neighboring. By identifying the similarity weight of users through their interaction history, a key concept of CF, we endeavor to build a user-user weighted connection graph based on their similarity weight.

In this study, we propose a recommendation framework, CombiGCN, in which item embeddings are only linearly propagated on the user-item interaction graph, while user embeddings are propagated simultaneously on both the user-user weighted connection graph and user-item interaction graph graphs with Light Graph Convolution (LGC) and combined in a simpler method by using the weighted sum of the embeddings for each layer. We also conducted experiments comparing CombiGCN with several state-of-the-art models on three real-world datasets.
\end{abstract}
\section{Introduction}
Recommendation systems play an important role in online businesses because of the economic benefits they bring by suggesting suitable products or services to customers. That motivation has driven research to improve algorithms to offer powerful recommendation engines, typically collaborative filtering (CF). Concurrent with the rise of deep learning, especially the use of GNNs to learn representations of users and items (as known as embeddings), many recent studies have focused on enriching embeddings by encoding them with collaborative signals, which carry information about user-item interactions \cite{ngcf, wigcn, lightgcn, sociallgn, sept}. These signals are extracted through the message-passing architecture of GNNs. More specifically, considering a user $u$ as a node in the graph whose embedding is $e_u$, at each propagation time this user node will adjust its embedding by aggregating all embeddings of neighboring items. During the aggregation progress, each embedding $e_i$ from a neighboring node item $i$ will be multiplied by a coefficient $p_{ui}=1/\sqrt{|\mathcal{N}_u||\mathcal{N}_i|}$, where $\mathcal{N}_u$ and $\mathcal{N}_i$ denote the first-hop neighbors of user $u$ and item $i$, so the updated embedding value $e_u$ not only carries information about neighboring items, it also reflects the mutual importance between user $u$ and item $i$ through the $p_{ui}$ coefficient.
%%%%Should shorten this paragraph.
%%Remove: Despite the contribution of the $p_{ui}$ coefficient, this coefficient is only reflected by the sum of the number of neighbor uesrs $\mathcal{N}_i$ of item $i$ and the number of neighbor items $\mathcal{N}_u$ of user $u$ during the embedded update process of user $u$ and item $i$ without showing the degree of connection.

However, GNN models only help each user or item embedding in the user-item interaction graph to be similar to neighboring nodes without regard to the weights of the links between nodes during the entire propagation process. There has been some research on adding weights to the embedding encoding process, such as \cite{wigcn, sept}. These studies construct a user-user graph where each connection between two users is the number of items shared by them. Aiming at addressing this problem, we have normalized the user-user graph based on Jaccard similarity and integrated these weights, which improves the quality of the extracted collaborative signal over each propagation and produces satisfactory embeddings for CF. Combining embedding propagated from two graphs has also been conducted through many studies \cite{sept, sociallgn, wigcn}. SocialLGN has proven that their proposed graph fusion operation is a state-of-the-art combination embeddings from the two graph method \cite{sociallgn}. Their results are more accurate than results from graph fusion operations based on GCN methods \cite{gcn} or GraphSage \cite{graphsage}. %During our research, we concluded that using too many trainable transformation matrices and activation functions in the graph fusion operation of SocialLGN makes this operation complex and inefficient. The model LightGCN \cite{lightgcn} also shows a negative impact of the trainable transformation matrices and non-linear activation function on problems with little or no semantics, and the collaborative filtering progress uses only user identifiers (IDs) and items to initialize embeddings.

In this paper, we propose a model named CombiGCN based on Light Graph Convolution (LGC) \cite{lightgcn} to propagate user and item embeddings on the user-item interaction graph; in the meantime, user embedding is also propagated on the user-user weighted connection graph. To fuse two user embeddings obtained after each propagation into an integrated embedding, instead of using the fusion graph operation of SocialLGN, we simply use the weighted sum of the embeddings. We demonstrate the superior performance of CombiGCN by comparing it with state-of-the-art recommendation models on three real-world datasets that we preprocessed to avoid cold-start and over-fitting.

\import{./}{background.tex}
\import{./}{method.tex}

\input{experiments}
\input{conlusion}

\end{document}

%% file: background.tex
\section{Related Work}

\subsection{Graph Convolution Networks}
Due to its superior ability to model graph-structured data, Graph Neural Networks (GNNs) have become the state-of-the-art technology for recommendation systems. A graph convolution network (GCN) is a special type of GNNs that uses convolution aggregations. Spatial GCNs based on 1-hop neighbor propagation and in each layer GCN neighborhood aggregations are required and thus the computational cost is significantly reduced. In the recommended context, spatial GCN contains NGCF \cite{ngcf}, WiGCN \cite{wigcn} and GCMC \cite{gcmc}. A recent study \cite{lightgcn} developed Light Graph Convolution (LGC) based on the GCN architecture but eliminates trainable matrices and non-linear activation functions in GCN. The experiment reported in \cite{lightgcn} also shows that LGC outperform GCN. Today's most modern CF models \cite{lightgcn, sociallgn, sept} also use LGC instead of traditional GCN.

\subsection{Multi-layer Perceptron and Backpropagation}
In machine learning models involving Neural Networks like GNNs, learning the trainable parameters includes forward-propagation and back-propagation. The forward-propagation stage calculates the embedding value of each node in the neural network with trainable parameters. In matrix form, these parameters include the trainable matrices $W^k$ and embeddings $E^k$ in the $k$-th layer. In addition, non-linear activation functions such as $ReLU$ and $tanh$ are also applied to the results. The calculations in forward-propagation produce the prediction result in the last layer. To train the model, an optimization function  to this result and propagate back to adjust the trainable parameters [$E, W$]. When training neural networks, forward-propagation and back-propagation depend on each other. For the recommendation problem using GNNs also follows this rule. However, a recent study \cite{lightgcn} has demonstrated the redundancy of the trainable matrices $W^k$ and non-linear activation functions in recommendation models, the reason being that the embeddings mapped from user and item IDs are not many features, so using too many trainable parameters makes the model heavy and ineffective.

%% file: method.tex
\section{Our proposed method}
We proposed our model, which includes a method to pre-process data set and the CombiGCN model, in this chapter. CombiGCN will explores the user and item interaction and weighted similarity matrix as the input, and make prediction at the output as recommendations. 
The overview of CombiGCN model is illustrated in Figure \ref{fig:overallmodel}.

\subsection{Pre-processing data}

\begin{algorithm}
\caption{Inference for data pre-processing }\label{alg:cap}
\begin{algorithmic}
\Require 
$\mathcal{U} \times \mathcal{I}$\Comment{Interaction between users and items in original dataset}

$r$\Comment{ratio between users and items you want to obtain}
\Ensure $R = U \times I  \subseteq \mathcal{U} \times \mathcal{I}$
\ForEach {item $i \in \mathcal I $}
    \State $ set_i \gets \textrm{list of users that interact with item } i$
    \State $len_i = \textrm{total number of users in }set_i$
\EndFor
\State $\widetilde{m} \gets \textrm{cardinality of set }I$
\State $I_{\textrm{c}} \gets \textrm{m items have highest } len_i$
\State $\widetilde{n} \gets \widetilde{m} \times r $
\ForEach {user $u \in \mathcal U $}
    \State $set_u \gets \textrm{list of items interacted by user } u$
    \State $sim_u = \textrm{Jaccard distance between} I_{\textrm{c}}  \textrm{and }set_u$
\EndFor
\State $U_{\textrm{c}} \gets \widetilde{n} \textrm{ users have highest }sim_u$
\State $U \gets n \textrm{ users from }\widetilde{n} \textrm{ users of set }U_{\textrm{c}} \textrm{ have more than 10 interactions with items}$
\State $I \gets m \textrm{ items have interacted by users in set }U$
\State \Return $R = U \times I$
\end{algorithmic}
\end{algorithm}

Algorithm \ref{alg:cap} brings two main benefits in the learning process of the recommendation model: \textbf{1) Avoid over fitting}, the dataset obtained from Algorithm has the most common features in the original dataset, so there will be little information about the typical interactions that cause the over fitting; \textbf{2)Remove noisy-negative interactions in implicit feedback}\cite{Gao_2022}, in the first step we have determined the set of items $I_{c}$ and throughout the next steps when collecting users we only collect interactions with items contained in set $I_{c}$. This will limit unpopular interactions, which are likely to be noisy-negative interactions.

\subsection{Adjacent graph and Weighted references matrix}
In this article, we use two graphs as the data sources including the user-item interaction graph and the user-user weighted connection graph denote by $G_R$ and $G_W$, $U = [u_1, \dotso , u_n] (|U| = n)$ denotes the user nodes across both $G_R$ and $ G_W$ and and $I = [i_1, \dotso , i_m] (|I| = m)$ denotes the item nodes in $G_R$. $R \in \mathbb{R}^{n \times m}$ is the binary matrix with entries only 0 and 1 that represent user-item interactions in $G_R$.

In WiGCN \cite{wigcn}, matrix $W_u = RR^T \in \mathbb{R}^{n \times n}$ accumulates the paths connecting two user via shared items. However, the matrix $W_u$ only shows the number of intersections between the two sets of items $I_i$ of user $u_i$ and $I_j$ of user $u_j$, but has not recorded the influence of couple of users {$i, j$} to all interaction data. We build the weight users matrix $W$ to represent the user-user weighted connection graph through the matrix $W_u$.
    \begin{equation}\label{eq:cal_W}
         W = W_u \varodot (D_{R}J + (D_{R}J)^T - W_u)^{-1}
    \end{equation}
Where, $D_R \in \mathbb{R}^{n \times n}$ is a diagonal matrix with each entry $D_{R_{ii}}$ represents the number of neighboring items of user $i$, $J \in \mathbb{R}^{n \times n}$ is the matrix of ones (or all-ones matrix) and $\varodot$ denote element-wise product. From a mathematical perspective,  each element of $W_u$ represents the intersection while $(D_{R}J + (D_{R}J)^T - W_u)$ represents the union of two list items of two users $u_i$ and $u_j$. Therefore Equation \ref{eq:cal_W} calculates the similarity between the pair of users {$i, j$} based on Jaccard Similarity. To avoid over-fitting the model when using both matrices $W$ and $R$ during the propagation process, we mapping the values of $W$ 
 to a number of discrete values in the interval $[0,1]$ where value $0.0$ represents no correlation between these two users while value 1.0 represents very high correlation. 

\subsection{CombiGCN}
\begin{figure}[h]
    \centering
    \includegraphics[scale=0.095]{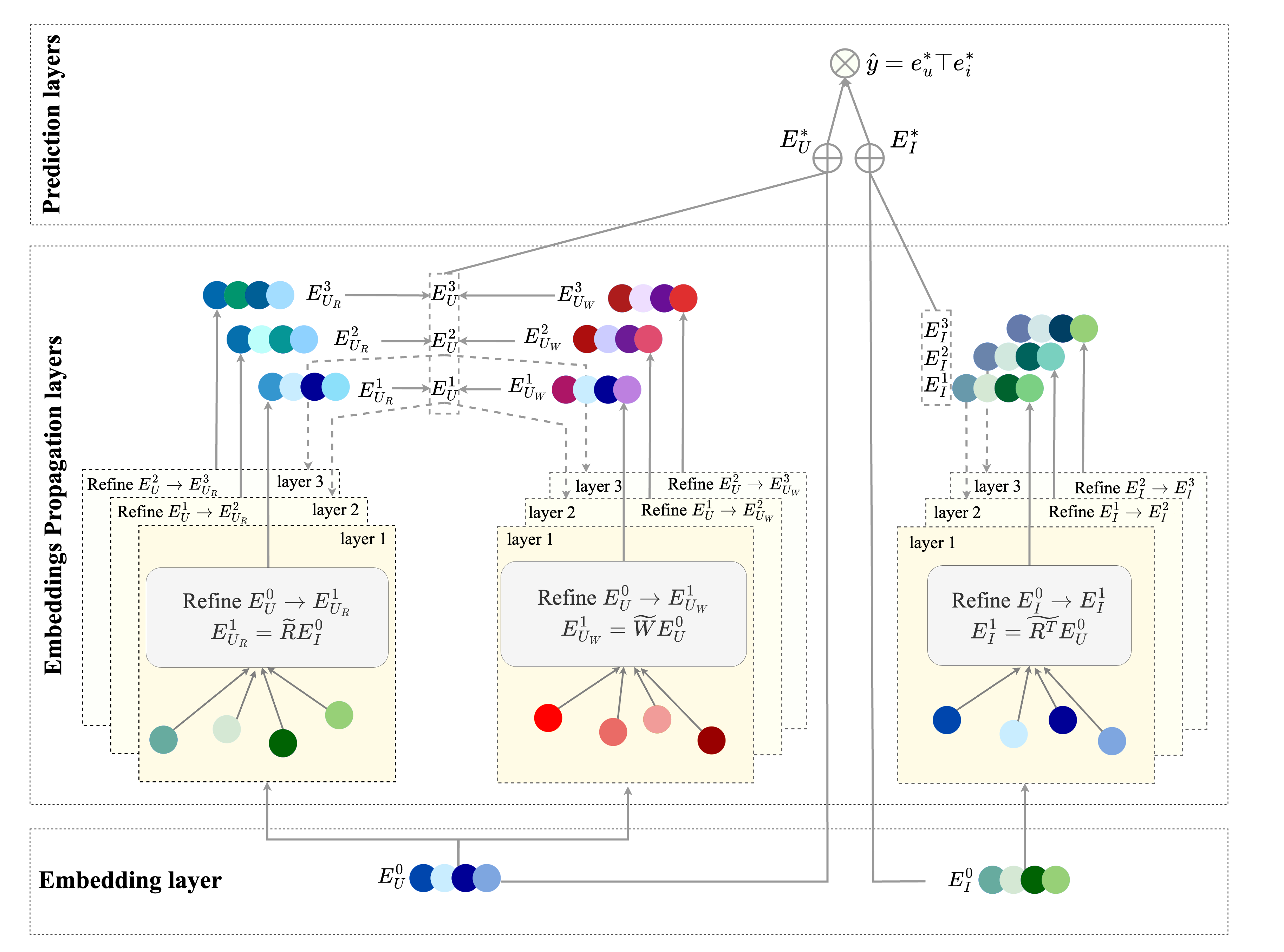}
    \caption{The architecture of the CombiGCN model}
    \label{fig:overallmodel}
\end{figure}

The general design of the proposed model is shown in Figure \ref{fig:overallmodel}, our model including three components - \textbf{1) Embeddings layer} that use the unique identifiers of users and items to create embedding,  \textbf{2) Propagation layers}, which propagate the representations of users and items in LGC architecture and,  \textbf{3) Prediction layer}, that predicts the score between users and items pair based on final embeddings obtained after $L$ propagation layers.
\subsubsection{Embedding layer}
Following the mainstream well-known models \cite{ngcf, wigcn, lightgcn}, we initialize user and item embeddings by map unique its ID into the latent space, and obtain dense vectors $e_u^{l=0}\in{\mathbb{R}^d} (e_i^{l=0}\in{\mathbb{R}^d})$. Where $l$ denote the number of layer propagation. The dimension of embeddings is denoted by $d$. We denote $E^l\in{\mathbb{R}^{(n+m) \times d}}$ is the set of all embeddings during propagation, i.e. $E^l$ contains the set of $n$ user embeddings  and $m$ item embeddings at $l$-th layer.
\begin{equation}
    E^l=E_U^l \parallel E_I^l=[e_{u_1}^l,\dotso ,e_{u_n}^l,e_{i_1}^l,\dotso,e_{i_m}^l]
\end{equation}
\subsubsection{Propagation layers}
In order to clearly introduce the embedding propagation process, we will first show this propagation process in the first layer of LGC architecture, and then show the general formula in the higher propagation layers. 
\paragraph{User embeddings propagation}
The input of first layer is embedding $E_U^0$, we will propagate this user embedding in two graphs, user-item interaction graph $G_R$ and user-user weighted connection graph $G_W$ respectively to obtain two user embeddings $E_{U_{R}}^1$ and $E_{U_{W}}^1$.
\begin{equation}
    E_{U_{R}}^1 = \widetilde{R}E_I^0 ;  E_{U_{W}}^1 = \widetilde{W}E_U^0
\end{equation}

We further define $\widetilde{R} = D_{R}^{-1/2}RD_{R^T}^{-1/2}$, where $D_R \in \mathbb{R}^{n \times n}$ is a diagonal matrix with each entry $D_{R_{ii}}$ represents the number of neighboring items of user $i$ and $D_{R^T} \in \mathbb{R}^{m \times m}$ is a diagonal matrix with each entry $D_{R^T_{jj}}$ represents the number of neighboring users of item $j$. Similarly, $\widetilde{W}$ is a symmetrically normalized matrix of $W$ and $\widetilde{W} = D_{W}^{-1/2}WD_{W}^{-1/2}$. We then combine the two embedding users $E_{U_{R}}^1$ and $E_{U_{W}}^1$ into $E_U^1$.
\begin{equation}
    E_U^1 = E_{U_{R}}^1 + E_{U_{W}}^1
\end{equation}

\paragraph{Item embeddings propagation}
For item embeddings, we just propagate them on LGC architecture only with user-item interaction graph. We also define $\widetilde{R^T} = D_{R^T}^{-1/2}R^TD_{R}^{-1/2}$.
\begin{equation}
    E_I^1 = \widetilde{R^T}E_U^0
\end{equation}

\paragraph{The general equation embeddings propagation}
We have presented the first propagation step in LGC architecture, in the next steps the process is similar, but the input will be user embeddings of the previous layer and not $E_U^0$ and $E_I^0$. Equation (8) represents the propagation processes of embedding at higher levels.
\begin{equation}
\begin{split}
    E^{l} = (E_{U_{R}}^l + E_{U_{W}}^l) \parallel E_I^l
    = (\widetilde{R}E_I^{l-1} + \widetilde{W}E_U^{l-1}) \parallel \widetilde{R^T}E_U^{l-1}
\end{split}\label{eq:Enext}
\end{equation}
\subsubsection{Prediction and optimization}
After $L$ embedding propagation layer we will get $L + 1$ embeddings, the arrival of $L + 1$ after $L$ propagation layer is due to including initial embedding $E^0$.
\begin{equation}
    E^*= \alpha_0E^0  + \alpha_1E^1 + \dotso + \alpha_LE^L 
\end{equation}

Where $\alpha_l =  1/(L  + 1)$, that mentioned in \cite{lightgcn} denotes the importance of the $l$-th layer embedding in constituting the final embedding. To perform model prediction, we conduct the inner product to estimate user preference for the target. 
\begin{equation}
    \hat{y}_{ui}= {e_u^*}^Te_i^*
\end{equation}
To learn parameters $\Phi = [E_U^0, E_I^0]$, CombiGCN have been applied Bayesian Personalized Ranking (BPR) \cite{bprmf}. BPR assumes observed interactions have higher preferences than an unobserved interactions. To optimize the prediction model we use mini-batch Adam \cite{adam}  and minimize the  BPR loss.
\begin{equation}
    Loss_{bpr} =  \sum_{\Omega^+_{ui}}\sum_{\Omega^-_{uj}} -ln \sigma ( \widehat{y}_{ui} - \widehat{y}_{uj} ) +  \lambda  \parallel  \Phi   \parallel ^2_2
\end{equation}

%% file: experiments.tex
\section{Experiments}
\subsection{Datasets Description}

We make experiments with our proposed model on three well-known datasets, which are Ciao, Epinions, and Foursquare. Each dataset is being pre-processed and divided into two sets: 80\% for training and 20\% for testing. %Table \ref{tab:data} show statistical information about the dataset after preprocessing.
%\begin{table}
%\centering
%\caption{Statistic of the experiment datasets.}
%\label{tab:data}
%\begin{tabular}{|l|r|r|r|r|}
%\hline
%Dataset &  \#Users & \#Items & \#Relations & Density\\
%\hline
%Ciao& 5,785 & 108,651 & 283,034 & 0.00045\\
%Epinions & 1,497 & 17,898 & 25,191 & 0.00094\\
%Foursquare & 29,902 & 42,541 & 1,737,759 & 0.00137\\
%\hline
%\end{tabular}
%\end{table}

\begin{itemize}
    \item \textbf{Ciao} \cite{ciao1, ciao2}: The Ciao dataset is an online shopping dataset containing the ratings given by users on a larger number of items.
    \item \textbf{Epinions} \cite{ciao1, ciao2}: Epinions is a popular online consumer review website.
    \item \textbf{Foursquare} \cite{4square1,4square2}: The Foursquare dataset record check-in data for different cities in the world.
\end{itemize}

\subsection{Experimental Settings}
\subsubsection{Setting parameters}
To ensure that the experimental results are fair, we set the parameters to be the same across all models. Specifically, the learning rate is 0.001, the coefficient of L2 normalization is 0.00001, and the number of layers of LGC is 3, with each layer having an embedding size of 64. We also use the same early stop strategy as NGCF and LightGCN; specifically, if in 50 consecutive epochs, the recall at 20 on the test result does not increase, the model will be stopped.

\subsubsection{Baseline.} We use the same datasets and repeat the experiments on all the following baseline models to demonstrate the result:
\begin{itemize}
    \item \textbf{BPR-MF} \cite{bprmf} is matrix factorization optimized by the Bayesian personalized ranking (BPR) loss, which exploits the user-item direct interactions only as the target value of interaction function.
    \item \textbf{GCMC} \cite{gcmc} adopts GCN encoder to generate the representations for users and items, where only the first-order neighbors are considered. Hence one graph convolution layer, where the hidden dimension is set as the embedding size.
    \item \textbf{WiGCN} \cite{wigcn} is developed on top NGCF and add the connection between each user-user and item-item pair in the interaction graph by the number of shared items or users.
    \item \textbf{NGCF} \cite{ngcf} conducts propagation processes on embeddings with several iterations. The stacked embeddings on output contains high-order connectivity in interactions graph. The collaborative signal is encoded into the latent vectors, making the model more sufficient.
    \item \textbf{LightGCN} \cite{lightgcn} focus on the neighborhood aggregation component for collaborative filtering. This model uses linearly propagating to learn users and items embeddings for interaction graph.
\end{itemize}

\subsection{Experiment Results}
\begin{table}
\centering
\caption{Overall Performance Comparisons}\label{tabResult}
\begin{adjustbox}{width=\columnwidth,center}
\begin{tabular}{| c|c| c|c|c| c|c|c| c|c|}
\hline
 \textbf{Dataset} & \multicolumn{3}{c|}{\textbf{Ciao}} & \multicolumn{3}{c|}{\textbf{Epinions}} & \multicolumn{3}{c|}{\textbf{Foursquare}}\\
   & precision & recall & ndcg &  precision & recall & ndcg & precision &  recall & ndcg \\ 
\hline
BPR-MF          & 0.01047 & 0.03182 & 0.02221 & 0.00087 & 0.00613 & 0.00330 & 0.01923 & 0.02479 & 0.02721 \\
GCMC        & 0.01439 & 0.04785 & 0.03484 & 0.00097 & 0.00825 & 0.00460 & 0.02066 & 0.03102 & 0.03006\\
NGCF        & 0.01596 & 0.05170 & 0.03825 & 0.00120 & 0.00955 & 0.00506 & 0.02094 & 0.03177 & 0.03107\\
WiGCN       & 0.01606 & 0.05317 & 0.03985 & 0.00147 & 0.01088 & 0.00683 & 0.02424 & 0.03433 & 0.03592\\
LightGCN    & 0.01673 & 0.05674 & 0.04294 & 0.00184 & 0.01219 & 0.00663 & 0.02612 & 0.03602 & 0.03778\\
\hline
%\hline
%\textbf{CombiGCN+fn}   & 0.01598 & 0.05014 & 0.03967 & 0.00164 & 0.01124 & 0.00658 & 0.02184 & 0.03634 & 0.03451\\
\textbf{CombiGCN}   & 0.01730 & 0.05845 & 0.04406 & 0.00204 & 0.01398 & 0.00720 & 0.02621 & 0.03801 & 0.03818\\
\hline
\end{tabular}
\end{adjustbox}
\label{figresult}
\end{table}
The overall performance comparison is shown in Table \ref{tabResult}. The results clearly show that our model consistently achieves the best performance in all three metrics and all three datasets. Further, MF performance is much inferior to that of GNN-based models because it cannot capture collaborative signals. Although GCMC uses GCN, it only captures neighborhood information in the first layer, so it is less effective than the NGCF and WiGCN. WiGCN has better accuracy than NGCF because it introduces information about the weights of users and items during embedding propagation, which makes the WiGCN model more efficient in capturing collaborative signals. LightGCN is an LGC-based model that has removed components that have been shown to negatively affect the model training process, so the results of LightGCN are very good, only worse than those of CombiGCN.

%We also discuss a variation of CombiGCN, CombiGCN+fn. The difference of this variant from our main model lies in the way the two user embeddings are combined two user embeddings . For CombiGCN+fn, we deploy a combination of two embeddings as equation (12), this is one of the embedding combine methods that brings good results and has been proven to be better than GCN \cite{gcn} and GraphSage \cite{graphsage} through research in SocialGCN\cite{sociallgn}.
%\begin{equation}
%    E_U^l = h^l / \parallel h^l \parallel _2 \text{ and } h^l = (W_3(\sigma(W_1\widetilde{R}E_I^{l-1} ) \parallel(W_2\widetilde{W}E_U^{l-1} )))
%\end{equation}

%During the research process, we found that when removing both the learnable parameters [$W_1, W_2, W_3$] and non-linear activation functions $\sigma$, combine embedding efficiency will be much better. This has also been proven through research \cite{yu2022graph}. CombiGCN only combines two embedding users by using the weighted sum of the embeddings for each layer as Equation \ref{eq:Enext}. This has also been proven through the results in Table \ref{tabResult}.
 

%% file: conlusion.tex
\section{Conclusion}
In this work, we attempted to improve the embedding quality by adding connection weights between users based on their interaction history. To do that, we introduce the CombiGCN model, which implements embedded functions on two graphs: a user-item interaction graph and a user-item weighted connection graph based on the Light Graph Convolution architecture. The key to CombiGCN lies in its ability to combine embedding functionality across multiple graphs in a simple and effective way. We provide three preprocessed datasets with our algorithm to reduce cold start, over-fitting, and data noise problems for evaluation experiments. The results of experiments with state-of-the-art models are a valuable demonstration of the success of weight addition and the multi-graph combination architecture of CombiGCN.